\documentclass[letterpaper,10pt,conference]{ieeeconf}

\IEEEoverridecommandlockouts
\usepackage{amsmath,amssymb,amsfonts}
\usepackage{graphicx}
\usepackage{booktabs}
\usepackage{multirow}
\usepackage{cite}
\usepackage{url}
\usepackage{xcolor}
\usepackage{tikz}
\usepackage{bm}
\usepackage{comment}
\usepackage{algorithm}
\usepackage{algpseudocode}

\makeatletter
\newcommand{\algmargin}{\the\ALG@thistlm}
\makeatother

\usepackage{mathtools}

\usepackage{hyperref}

\usetikzlibrary{
    arrows.meta,
    positioning,
    fit,
    calc,
    shapes.multipart
}

\title{\LARGE \bf Diffusion-Based Multiple-Shooting Indirect Optimal Control for Fuel-Optimal Spacecraft Trajectory Generation}

\author{
Saeid Tafazzol, Ehsan Taheri, and Ryne Beeson%
\thanks{Saeid Tafazzol and Ehsan Taheri are with the Department of Aerospace Engineering, Auburn University; \{szt0097,ezt0028\}@auburn.edu}%
\thanks{Ryne Beeson is with the Department of Mechanical and Aerospace Engineering, Princeton University; ryne@princeton.edu}%
\thanks{This work has been submitted to the IEEE for possible publication. Copyright may be transferred without notice, after which this version may no longer be accessible.}%
}

\begin{document}

\maketitle
\thispagestyle{empty}
\pagestyle{empty}

\begin{abstract}
Diffusion-based generative models (DMs) have found applications in control problems, and in particular robotics, where the DMs enable exploration of possible control solutions. 
A critical shortcoming of these applications is that they have lacked optimality guarantees.
This is a problem 
for their potential use in fuel-optimal spacecraft trajectories that are characterized with long time-horizons and bang-bang profiles. 
Alternatively, indirect optimal control methods ensure explicit satisfaction of necessary conditions, but are highly sensitive to the initial costate estimation needed to solve the resulting Hamiltonian boundary-value problems (HBVPs). 
To alleviate this sensitivity and enlarge the convergence domain of HBVPs, advanced indirect methods have been developed that use smoothing approaches and continuation. 
We propose a diffusion-based multiple shooting indirect control method that combines the exploration capability of DMs with indirect method to generate fuel-optimal spacecraft trajectories. 
We benchmark our method against an advanced indirect method on a fuel-optimal Earth-Mars low-thrust transfer problem, showing higher convergence robustness than the advanced indirect method that is based on random costate initialization. Code and visualizations are available at \href{https://saeidtafazzol.github.io/Diffusion_Indirect_Control/}{https://saeidtafazzol.github.io/Diffusion\_Indirect\_Control/}.

\end{abstract}
\vspace{-2mm}
\section{Introduction and related work}

Diffusion models (DMs) are a type of generative model trained to produce samples from a data distribution.
They have recently been used for policy learning, behavior generation, and trajectory generation \cite{janner2022planning,chi2024diffusionpolicy}. 
In the control setting, these methods have primarily been applied to imitation learning and behavior cloning; learning to reproduce expert demonstrations. 
While effective in these settings, the standard DM does not explicitly provide any guarantees on the optimality of sample solutions. 
To improve performance or diversity, many approaches incorporate guidance during the reverse diffusion process, including classifier guidance \cite{dhariwal2021diffusion}, classifier-free guidance \cite{ho2021classifier}, and task-specific value or policy guidance \cite{janner2022planning,jackson2024policy}. 
Although such guidance has been used to steer the generated trajectories toward lower-cost solutions, explicitly guiding solutions to satisfy necessary conditions for optimality has been underexplored.


Diffusion models have been trained on optimal costate data for low-thrust spacecraft trajectory problems \cite{Graebner:2025.jas.72.6, Graebner:2026.aiaa.scitech, Graebner:2026.jass.tba}, with the ability to produce costate samples for new and related problems that converge quickly and reliably. 
Unlike the aforementioned literature, this work generates joint state--costate trajectory samples. 
This work also makes the novel contribution of providing context to a transformer-based DM in the form of the integrated state--costate trajectories with a one-step lag. 
This feature is hypothesized to provide implicit guidance to the diffusion process and encourage samples that nearly satisfy the necessary conditions for optimality. 

Several studies have used other forms of deep learning to warm-start indirect trajectory optimization \cite{chen2023fast,shi2022neural,yin2020low}. 
Unlike this work, these studies used traditional machine learning and not a generative model. 
This means that the networks in the cited works are deterministic and therefore may be fragile in more complex control problems, which benefit from trialing several good initial guesses.  
Ref.~\cite{lian2025co} is similar to the present paper in that it predicts costates throughout the trajectory.
Yet it differs in that it only predicts the costates; the network is deterministic and not as capable as the transformer-based approach used here for high-dimensional sequential data; the authors only test on a one-dimensional problem \cite{lian2025co}, whereas here we are considering problems with  7 states and 4 controls 
and apply a multiple-shooting transcription. 

The indirect method \cite{taheri2023l2} guarantees that solutions satisfy necessary conditions, but is challenged by unknown costates that are numerically sensitive \cite{betts1998survey}. 
In this paper a novel complementary approach of DMs and the indirect approach are introduced. 
The DM is used to provide diverse initial guesses for the indirect method, exploring the admissible solution space and alleviating some difficulties in guessing the unknown costates. 
And in a complementary fashion, the indirect method provides implicit guidance to the transformer based DM via context learning, which helps to guide samples toward satisfaction of the necessary conditions. 

To re-iterate, the main contributions of this work include: 
1) The first work (to the best of our knowledge) to introduce implicit one-step lag guidance to a diffusion model with multiple-shooting indirect methods to produce joint state--costate trajectories subject to boundary conditions; 
2) The use of a transformer-based architecture that handles the high-dimensional sequential representation and naturally allows the context learning from the one-step lag guidance; 
3) A novel modification to efficiently generate a diverse training dataset that satisfies the necessary conditions \cite{izzo2021real}.

\section{Methodology}
\label{sec:methodology}

\subsection{Background on indirect optimal control}
Indirect methods produce extremal trajectories by solving for state and costate variables that satisfy the Hamiltonian dynamics, boundary conditions, (any) transversality conditions, and in a multiple-shooting method, continuity between neighboring segments. Consider the optimal control problem
(OCP):
\vspace{-3mm}
\vspace*{-4mm}
\begin{equation} \label{eq:generic_ocp}
\begin{aligned}
&\min_{\bm{x} \in \mathcal{X},\bm{u} \in \mathcal{U}} \mathcal{J} = \int_{t_0}^{t_f} L(\bm{x}(t),\bm{u}(t),t)\,dt,\\
&\text{subject to},\\
&\dot{\bm{x}}(t) = \bm{f}(\bm{x}(t),\bm{u}(t),t), \text{(state dynamics)}\\
& \bm{x}(t_0) = \bm{x}_0,\bm{\psi}(\bm{x}(t_f)) = \bm{0},\text{(terminal conditions),}
\end{aligned}
\end{equation}
where $\bm{x} \in  \mathbb{R}^{n_x}$ and $\bm{u}\in  \mathbb{R}^{n_u}$ are the states and controls, with $\mathcal{X}$, $\mathcal{U}$ their admissible sets, respectively. 
Let $\bm{\lambda} \in \mathbb{R}^{n_x}$ be the costate vector. 
Then the control Hamiltonian is
\begin{equation} \label{eq:generic_hamiltonian}
H(\bm{x},\bm{\lambda},\bm{u},t)
=
L(\bm{x},\bm{u},t)
+
\bm{\lambda}^\top
\bm{f}(\bm{x},\bm{u},t),
\end{equation}
and the costate dynamics and extremal control derived using Pontryagin's minimum principle (PMP) are
\begin{align}
\dot{\bm{\lambda}}^\top(t)
&=
-\frac{\partial H}{\partial \bm{x}} = \bm{g}^\top(\bm{x},\bm{\lambda},\bm{u},t),
\label{eq:generic_costate}
\\
\bm{u}^{*}(t)
&=
\arg\min_{\bm{u} \in \mathcal{U}}
H(\bm{x},\bm{\lambda},\bm{u},t).
\label{eq:generic_optimal_control}
\end{align}

Eqs.~\eqref{eq:generic_costate}--\eqref{eq:generic_optimal_control} and state dynamics along with the state boundary conditions and (any) transversality conditions form the optimality conditions (superscript $^{*}$ denotes an extremal control). 
The necessary conditions result in an HBVP that is solved numerically for all practical OCPs. 
Defining the augmented state--costate vector as $\bm{z}^\top
    =
    \begin{bmatrix}
        \bm{x}^\top,
        \bm{\lambda}^\top
    \end{bmatrix}$,

In fuel-optimal spacecraft trajectory problems, bang-bang control profiles arise from PMP and introduce discontinuities in the dynamics that degrade the convergence performance of gradient-based solvers.
To alleviate this issue, smoothing or regularization procedures \cite{taheri2018generic,taheri2023l2,tafazzol2024comparison} are often invoked. 

Two typical solution approaches are \textit{single-shooting} and \textit{multiple-shooting}. In a conventional single-shooting formulation, a model would predict only the initial costate $\bm{\lambda}(t_0)$ and rely on forward integration to produce the complete trajectory. This is a low-dimensional prediction problem, but it is also a sensitive one, where a small change in $\bm{\lambda}(t_0)$ may lead to a dramatically different terminal state. Multiple shooting \cite{nurre2025constrained} breaks the problem into multiple smaller-duration segments. Although this is a higher-dimensional prediction problem, it provides a substantially richer learning signal and gives the model considerably more control over the otherwise numerically sensitive process of costate recovery and aligns naturally with diffusion-based trajectory generation.

The time is discretized as
$t_0<t_1<\cdots<t_{N-1}$, where
$\bm{z}_i\coloneq\bm{z}(t_i)$. Furthermore, let
\begin{equation}
    \dot{\bm{z}}(t)
    =
    \bm{F}(\bm{z}(t),t)
    \coloneq
    \begin{bmatrix}
        \bm{f}(\bm{x},\bm{u}^*,t)\\[1mm]
        \bm{g}(\bm{x},\bm{u}^*,\bm{\lambda},t)
    \end{bmatrix},
\end{equation}
denote the coupled Hamiltonian state--costate dynamics. For each shooting interval $[t_i,t_{i+1}]$, the states and costates are propagated forward in time. An admissible multiple-shooting extremal must therefore satisfy
\begin{equation}
\label{eq:generic_multiple_shooting_problem}
\begin{aligned}
    &\bm{x}_0 = \bm{x}_{\mathrm{init}},~
    \bm{\psi}(\bm{z}_{N-1}) = \bm{0},\\
    &\bm{z}_{i+1}
    =
    \bm{z}_i
    +
    \int_{t_i}^{t_{i+1}}
    \bm{F}(\bm{z}(t),t)\,dt,~ i=0,\ldots,N-2.
\end{aligned}
\end{equation}

\vspace*{1mm}
The final relation in Eq.~\eqref{eq:generic_multiple_shooting_problem} enforces continuity between adjacent shooting segments. 
The discrete trajectory is collected into the node matrix
\begin{equation}
    \begin{bmatrix}
        \bm{z}_0,
        \bm{z}_1,
        \cdots,
        \bm{z}_{N-1}
    \end{bmatrix}
    \in\mathbb{R}^{N\times n_z},
\end{equation}
where $n_z=2n_x$. From this point onward, the column-wise vectorized form of the complete trajectory is denoted by

\begin{equation} \label{eq:big_vec_definition}
    \bm{Z}
    \coloneq
    \operatorname{vec}\!\left(\begin{bmatrix}
        \bm{z}_0,
        \bm{z}_1,
        \cdots,
        \bm{z}_{N-1}
    \end{bmatrix}\right)
    \in
    \mathbb{R}^{Nn_z}.
\end{equation}

To distinguish between trajectory and diffusion times, $\bm{Z}^{(j)}$ denotes the complete vectorized trajectory at diffusion step $j$, whereas $\bm{z}_i$ is the state--costate value at shooting node $i$.

\subsection{Diffusion-based trajectory generation}

We make use of a specific type of diffusion modeling formulation known as the denoising diffusion probabilistic model (DDPM) \cite{sohl2015deep}. 
It consists of two complementary stochastic processes: a forward process and a reverse denoising one. 
The forward process is chosen as a Markov chain with the ultimate aim of transforming the general data distribution to an easily sampled distribution, such as a Gaussian. 
This action is achieved by running each training data sample forward under the prescribed Markov chain. 
A generative model is then trained to learn a close approximation to the reverse process, which can be used at inference to generate new samples that ``explore" the space of the original data distribution. 

Formally, the DDPM in this case operates as follows. Let $\bm{Z}^{(0)}$ denote a ``clean'' trajectory drawn from the data distribution, represented in this work by a fixed dataset whose construction is explained later on. The forward process, defined by a homogeneous Markov chain that is iterated for $J$-steps, is normally given by the transition density
\begin{equation}
q(\bm{Z}^{(j)}|\bm{Z}^{(j-1)})
=
\mathcal{N}
\left(
\sqrt{\alpha^{(j)}}\,\bm{Z}^{(j-1)},
(1-\alpha^{(j)})\mathbf{I}
\right),
\label{eq:forward_markov}
\end{equation}
where $\alpha^{(j)} = 1-\beta^{(j)}$ and $\{\beta^{(j)}\}_{j=1}^{J}$ are predefined noise schedules. 
Any initial data distribution will asymptotically tend to a standard multivariate Gaussian under this Markov chain, but in practice this is only carried out for sufficiently high $J$ steps. A key property of choosing a simple Markov chain is that the forward process admits a closed-form
\begin{equation}
q(\tilde{\bm{Z}}^{(j)}|\bm{Z}^{(0)})
=
\mathcal{N}
\left(
\sqrt{\bar{\alpha}^{(j)}}\,\bm{Z}^{(0)},
(1-\bar{\alpha}^{(j)})\mathbf{I}
\right),
\label{eq:forward_closed_form}
\end{equation}
where
\(
\bar{\alpha}^{(j)}
=
\prod_{i=1}^{j}\alpha^{(i)}.
\)
The closure of Gaussians under affine transformations yields the result
\begin{equation}
\tilde{\bm{Z}}^{(j)}
=
\sqrt{\bar{\alpha}^{(j)}}\,\bm{Z}^{(0)}
+
\sqrt{1-\bar{\alpha}^{(j)}}\,\bm{\epsilon},
\qquad
\bm{\epsilon}\sim\mathcal{N}(\bm{0},\mathbf{I}).
\label{eq:forward_sampling}
\end{equation}

Here, we used $\tilde{\bm{Z}}$ to differentiate between the ``corrupted'' trajectory that is normally used and what we use in this work. Here, to enforce the boundary conditions, we instead only corrupt the free variables along the trajectory and not the boundary conditions:

\begin{equation}
\label{eq:masked_noisy_trajectory}
    \bm{Z}^{(j)}
    =
    \bm{m}\odot\bm{Z}^{(0)}
    +
    (\bm{1}-\bm{m})\odot\widetilde{\bm{Z}}^{(j)},
\end{equation}
where $\bm{m}\in\{0,1\}^{Nn_z}$ is the componentwise condition mask.
\newpage
\vspace*{0mm}
{\linespread{0.9}\selectfont
The reverse process then begins with an initial sample from a standard multivariate Gaussian and after $J$-steps produces a new sample from the data distribution. 
The role of learning enters at this stage to approximate the reverse chain. 
In particular, the assumptions of the forward process mean that the reverse transition density is a Gaussian parameterized as
\par}
\vspace{-6mm}
\begin{equation}
p_{\theta}(\bm{Z}^{(j-1)} | \bm{Z}^{(j)})
=
\mathcal{N}
\left(
\bm{Z}^{(j-1)}; 
\bm{\mu}_{\theta}(\bm{Z}^{(j)},j),
\tilde{\beta}^{(j)}\mathbf{I}
\right),
\label{eq:reverse_distribution}
\end{equation}
with 
\begin{align}
\bm{\mu}_{\theta}
&=
\frac{
\sqrt{\bar{\alpha}^{(j-1)}}\beta^{(j)}}
{
1-\bar{\alpha}^{(j)}
}
\bar{\bm{Z}}^{(0)}_\theta
+
\frac{
\sqrt{\alpha^{(j)}}(1-\bar{\alpha}^{(j-1)})
}{
1-\bar{\alpha}^{(j)}
}
\bm{Z}^{(j)},
\label{eq:reverse_mean} \\
\bar{\bm{Z}}^{(0)}_\theta
&=\frac{
\bm{Z}^{(j)}
-
\sqrt{1-\bar{\alpha}^{(j)}}\,
\bm{\epsilon}_{\theta}(\bm{Z}^{(j)},j)
}{
\sqrt{\bar{\alpha}^{(j)}}
},
\label{eq:clean_sample_estimate}
\end{align}
and
\(
\tilde{\beta}^{(j)}
=
\frac{
1-\bar{\alpha}^{(j-1)}
}{
1-\bar{\alpha}^{(j)}
}
\beta^{(j)}.
\)
This is the standard $\epsilon$-prediction formulation \cite{ho2020denoising}, and we use a transformer-based neural network (NN) to learn $\bm{\epsilon}_\theta$. 
To reduce computational demands during inference, a reduced timestep schedule for the reverse process is commonly used, such as
\begin{equation}
\{j_1,j_2,\ldots,j_M\},
\quad
J=j_1 > j_2 > \cdots > j_M=0,
\end{equation}
with $M \ll J$. 
The reverse transitions are then performed only between consecutive elements of the reduced schedule,
\(p_{\theta}(\bm{Z}^{(j_{k+1})}|\bm{Z}^{(j_k)})
\).
Empirically, this can provide computational speed-up with only subtle degradation in some problems (see \cite{Elango:2026.aas.asc.1049} for an extensive study on DM formulations applied for low-thrust spacecraft). 
A rigorous treatment to handle the reverse process transitioning to a non-Markovian process is often ignored, and as an initial study, this more complex treatment is also not included in this work.

\subsection{Integrated State and Costate Signals}

A standard diffusion transformer could operate only on the noisy state and costate nodes. We instead provide the model with an explicit view of what the Hamiltonian dynamics predict between neighboring nodes. This is the central dynamics-informed component of the proposed architecture.

At diffusion step $j$, the noisy node values are written as
$\bm{z}^{(j)}_i=[(\bm{x}^{(j)}_i)^{\top},
(\bm{\lambda}^{(j)}_i)^{\top}]^{\top}$. For every segment
$i=0,\ldots,N-2$, the current noisy state and costate are integrated forward over $[t_{i}, t_{i+1}]$:
\begin{equation}
\label{eq:integrated_signal}
    \widehat{\bm{z}}^{(j)}_{i+1}
    =
    \bm{z}^{(j)}_i
    +
    \int_{t_i}^{t_{i+1}}
    \bm{F}(\bm{z}(t),t)\,dt
    =
    \begin{bmatrix}
        \widehat{\bm{x}}^{(j)}_{i+1}\\
        \widehat{\bm{\lambda}}^{(j)}_{i+1}
    \end{bmatrix}.
\end{equation}

The sequences
$\{\widehat{\bm{x}}^{(j)}_{i+1}\}_{i=0}^{N-2}$ and
$\{\widehat{\bm{\lambda}}^{(j)}_{i+1}\}_{i=0}^{N-2}$ are not used to overwrite the current trajectory. Instead, they are embedded as two additional token streams. The model can therefore observe both the independently proposed node
$\bm{z}^{(j)}_{i+1}$ and the value obtained by integrating its predecessor
$\widehat{\bm{z}}^{(j)}_{i+1}$.

This gives the transformer direct access to the information underlying the multiple-shooting residual
\begin{equation}
    \bm{r}^{(j)}_i
    =
    \bm{z}^{(j)}_{i+1}
    -
    \bm{z}^{(j)}_i
    -
    \int_{t_i}^{t_{i+1}}
    \bm{F}(\bm{z}^{(j)}(t),t)\,dt.
\end{equation}

\vspace*{0mm}
We do not explicitly place $\|\bm{r}^{(j)}_i\|$ in the training loss. Instead, the propagated nodes are supplied as context, allowing self-attention to learn how local dynamic inconsistency should influence the global denoising update. In this sense, the numerical integrator \textit{does not} solve the HBVP for the model; it merely shows the model where its present trajectory would go if the Hamiltonian dynamics were to be satisfied.

The integrated values are treated as auxiliary, non-differentiated signals. Gradients are not propagated through the numerical ordinary differential equation (ODE) solution. If integration fails on a segment, the unavailable integrated state and costate are replaced with learned failure tokens rather than silently presenting zeros as physically meaningful predictions. It can then technically be said that the model is learning $p_\theta(\bm{Z}^{(j_{k+1})}| \bm{Z}^{(j_k)} , \widehat{\bm{Z}}^{(j_k)})$ where $\widehat{\bm{Z}}^{(j_k)}$ is the trajectory with one step-lag.



\subsection{Transformer Architecture}

\begin{figure}[t]
\vspace{3mm}
\centering
\input{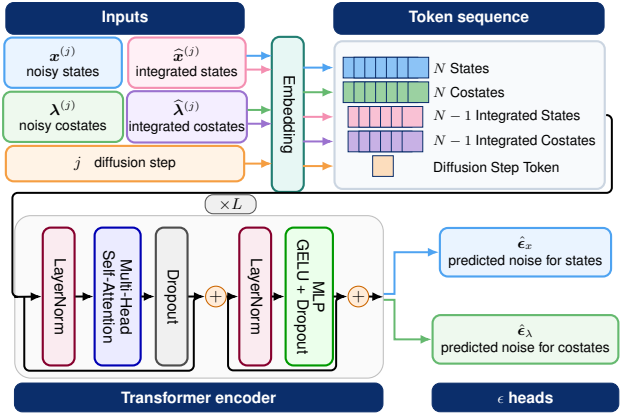}
\caption{transformer architecture used for the diffusion model.}
\label{fig:diffusion-indirect-control-transformer}
\vspace{-5mm}
\end{figure}

The complete architecture is shown in
Fig.~\ref{fig:diffusion-indirect-control-transformer}. At every diffusion step, the model receives five token streams:

\begin{enumerate}
    \item $N$ noisy state tokens;
    \item $N$ noisy costate tokens;
    \item $N-1$ forward-integrated state tokens;
    \item $N-1$ forward-integrated costate tokens; and
    \item one sinusoidal diffusion-step token.
\end{enumerate}

Each state and costate value is projected into the common embedding dimension using a learned linear map. Learned positional embeddings identify the associated shooting node, while learned condition embeddings indicate whether one or more components of that node are fixed by the boundary conditions. Integrated state and costate tokens use the positional embedding of the node they predict, namely node $i+1$ for the integration beginning at node $i$.

The concatenated token sequence is processed by $L$ pre-normalized transformer encoder blocks. Each block consists of multi-head self-attention, dropout, a residual connection, LayerNorm, and a feed-forward network with GELU activation. Following the final LayerNorm, hidden representations corresponding to the original state and costate tokens are passed through separate output heads to predict
$\hat{\bm{\epsilon}}_x$ and $\hat{\bm{\epsilon}}_{\lambda}$. It should be noted that while DM gets very close to an admissible trajectory, some residuals remain. Therefore, we use a NonLinear Program (NLP) solver for numerical refinement, initialized with the trajectory generated by the diffusion model.

\subsection{Admissible Trajectory Dataset}

As mentioned, a dataset of trajectories satisfying the Hamiltonian dynamics and the relevant endpoint constraints is required.
While there are methods for generating costates \cite{ayyanathan2022mapped}, an efficient strategy is to integrate backward from a condition that is common to all fuel-optimal trajectories,  $\lambda_m(t_f) =0$. Randomizing the final state $\bm{x}(t_f)$, and integrating backward correspond to an extremal solution that satisfies the first-order necessary conditions for fuel-optimality, and will correspond to an arbitrary initial state $\bm{x}(t_0)$.

The resulting dataset is written as
$\mathcal{D}=\{\bm{Z}_m\}_{m=1}^{M}$, where every
$\bm{Z}_{m}$ contains the state and costate values at the same $N$ shooting nodes. Additional problem-dependent rejection, normalization, and filtering steps may be necessary to remove singular, numerically failed, or otherwise undesirable trajectories. Since these steps depend on the specific dynamical system, they are detailed in experimental sections.

\section{Experiment: Earth-to-Mars Transfer}
\label{sec:e2m}

A fixed-time fuel-optimal Earth-to-Mars benchmark problem is considered. The spacecraft heliocentric trajectory is optimized using two-body dynamics assumptions. 
The state is defined as $\bm{x}^\top
=
\begin{bmatrix}
\bm r^\top
&
\bm v^\top
&
m
\end{bmatrix}^\top \in \mathbb{R}^{n_x = 7}$, where $\bm r \in \mathbb{R}^3$ and $\bm v \in \mathbb{R}^3$ denote the heliocentric position and velocity vectors, respectively, and $m$ is the spacecraft mass.

The spacecraft departs from Earth and arrives at Mars after a transfer duration of $348.795$ days. 
The mission parameters and boundary conditions are summarized as follows \cite{taheri2016enhanced}:
$\mu \approx 1.3271244 \times 10^{11} \; (\mathrm{km}^3/\mathrm{s}^2)$;
$m_0 = 1000 \; (\mathrm{kg})$;
$I_{\mathrm{sp}} = 2000 \; (\mathrm{s})$;
$T_{\max} = 0.5 \; (\mathrm{N})$. The position and velocity vectors for boundary conditions are taken from \cite{taheri2016enhanced}.

The control vector, $\bm{u}^\top = [\hat{\bm{\alpha}}^\top, \delta] \in \mathbb{R}^{n_u=4}$ consists of a thrust steering unit direction, $\hat{\bm \alpha} \in \mathbb{S}^2 \subset \mathbb{R}^3$, and a throttle input,  $\delta \in [0,1]$. The equations of motion are
\begin{align}
\dot{\bm r}
=
\bm v,~
\dot{\bm v}
=
-\frac{\mu_s}{r^3}\bm r
+
\frac{T_{\max}}{m}
\delta
\hat{\bm \alpha},~
\dot m
=
-\frac{T_{\max}}{c}\delta,
\end{align}
where $r = \|\bm{r}\|_2$, $\mu_s$ is the solar gravitational parameter, $T_{\max}$ is the maximum available thrust, and $c = I_{\mathrm{sp}} g_0$ is the effective exhaust velocity with $I_{\mathrm{sp}}$ denoting the specific impulse and $g_0$ being the sea-level gravity magnitude. The fuel-optimal cost functional is
\begin{equation}
\min_{\hat{\bm{\alpha}},\delta} J
=
\int_{t_0}^{t_f}
\frac{T_{\max}}{c}\,
\delta(t)\,dt.
\end{equation}

Let $\bm{\lambda}^\top = [\bm{\lambda}^\top_{\bm{r}},\bm{\lambda}^\top_{\bm{v}},\lambda_m]$ denote the costate vector. From Sec.~\ref{sec:methodology} and applying PMP yields the state--costate dynamics and optimal control laws.
The optimal thrust direction is given by $\hat{\bm\alpha}^{*} = -\bm\lambda_{\bm{v}}/\|\bm\lambda_{\bm{v}}\|_2$, while the throttle is determined by the switching function, $S$, as
\begin{equation}
\delta^* \in \begin{cases}
    1,~S>0,\\
    0,~S<0,
\end{cases}~\text{with}~S
=
\frac{c\|\bm\lambda_{\bm{v}}\|_2}{m}
+
\lambda_m
-
1.
\end{equation}

Thus, fuel-optimal solutions exhibit bang-bang thrust profiles. Trajectory generation and numerical integration were implemented in Python using \texttt{Diffrax} \cite{kidger2021on}, while symbolic dynamics construction and compilation were performed using \texttt{Jaxadi} \cite{jaxadi2024}. The multiple-shooting refinement stage was implemented using \texttt{CasADi} \cite{Andersson2019} with IPOPT \cite{wachter2006implementation}.

\subsection{Training Dataset Generation}

A final position vector is sampled within a bounded region of the normalized heliocentric state space. A dynamically admissible velocity magnitude is then computed using the vis-viva relation, $v=
(\mu
\left(
\frac{2}{r}
-
\frac{1}{a}
\right))^{0.5}$,
where $a$ is a randomly selected semi-major axis. The velocity direction is chosen randomly within the plane orthogonal to the sampled position vector. A terminal spacecraft mass is sampled independently, and the costate for mass at final time is set to zero $\lambda_m(t_f)=0$. while the remaining terminal costates are sampled from bounded distributions. The resulting terminal state--costate vector is then propagated backward using the Hamiltonian dynamics to generate a complete extremal trajectory. Although this procedure generates dynamically admissible trajectories, most samples exhibit an undesirable behavior. In particular, backward propagation frequently produces excessively large costate magnitudes near the beginning of the trajectory. Through the switching function, these large costates lead the throttle into $\delta(t)  = 1$, producing trajectories without bang-bang thrust profiles.

\begin{figure}[t]
    \vspace{2mm}
    \centering
    \includegraphics[width=0.872\columnwidth]{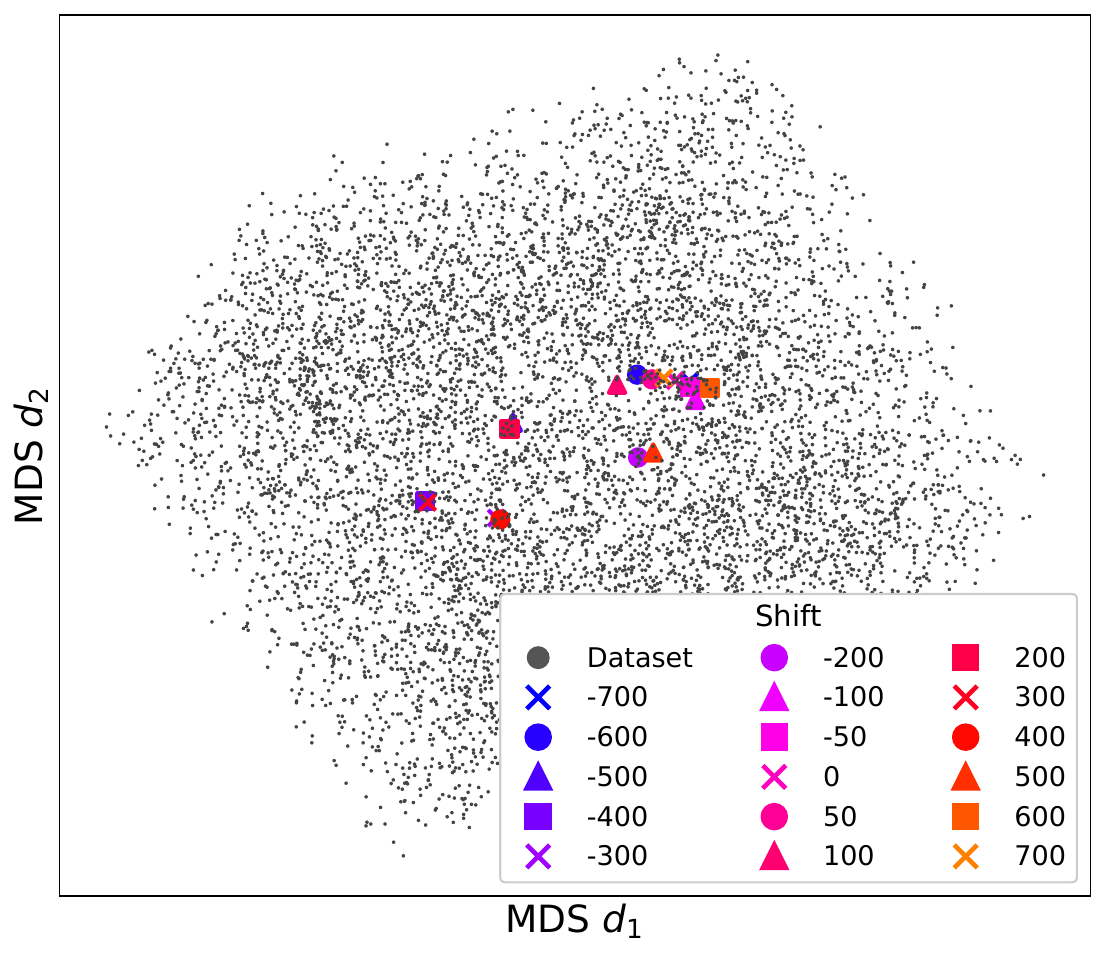}
    \caption{\textbf{Dataset Coverage:} To visualize the training dataset, the $14D$ effective input to the model for each trajectory was projected down to a 2D space using multidimensional scaling, which attempts to preserve the distances between trajectories in this 2D space.}
    \label{fig:projected_2d_dataset}
    \vspace{-4mm}
\end{figure}

To address this issue, a gradient-descent refinement procedure is applied to the sampled terminal conditions before a trajectory is accepted into the dataset. The optimization variables consist of the terminal position and velocity, along with their respective costates, while the terminal mass and terminal mass costate remain fixed. The objective is to reduce excessively large recovered costates and reject trajectories that produce unrealistic initial conditions after backward propagation. Empirically, this refinement step substantially improves dataset quality and yields trajectories exhibiting the expected bang-bang thrusting behavior; using this procedure, $49,152$ trajectories were generated.
To show their coverage one can look at Fig.~\ref{fig:projected_2d_dataset}; to plot this, only the masked states and costates that are given as fixed constraints to the model, namely $\bm{x}(t_0), m(t_0),\bm{x}(t_f),\lambda_m(t_f)$ which construct a $14D$-vector that is then projected down to a 2D space using multidimensional scaling \cite{kruskal1978multidimensional} which tries to preserve their respective distances in the projected space.

\begin{table}[htbp]
  \centering
  \caption{Model and inference parameters.}
  \label{tab:training_inference_parameters}
  \begin{tabular}{@{}ll@{\quad}ll@{}}
    \toprule
    \textbf{Parameter} & \textbf{Value}
    & \textbf{Parameter} & \textbf{Value} \\
    \midrule
    LR decay epochs (linear)     & 24
    & Batch size                 & 256 \\
    Forward diffusion levels $J$ & 5000
    & Epochs                     & 30 \\
    Denoising steps (inference)  & 30
    & Weight decay               & $10^{-6}$ \\
    Initial learning rate        & $3\times10^{-3}$
    & Dropout                    & 0.1 \\
    Final learning rate          & $10^{-5}$
    & Sequence length            & 32 \\
    IPOPT tolerance              & $10^{-15}$
    & Embedding dimension        & 512 \\
    Sample clip range            & $\pm 5.0$
    & Transformer layers         & 12 \\
                                 &
    & Attention heads            & 4 \\
                                 &
    & MLP ratio                  & 4 \\
    \bottomrule
  \end{tabular}
\end{table}




\subsection{Time-Shift Generalization Study}

We perform a time-shift study to evaluate the ability of the diffusion model to generalize beyond the nominal Earth-to-Mars transfer window. The Earth-to-Mars transfer is used only for evaluation (and not as the training dataset). The model and solver parameters can be found in Table~\ref{tab:training_inference_parameters}. The heliocentric states of Earth and Mars are propagated forward (or backward) in time under two-body dynamics, while keeping the transfer duration fixed at $348.795$ days. For each shifted transfer window, the complete pipeline is executed $250$ times using independent random seeds. Each trial begins from a newly sampled Gaussian latent trajectory and therefore produces a different diffusion-generated initialization before the multiple-shooting refinement stage. As a baseline, a variant of the works \cite{taheri2023l2,tafazzol2024comparison} was used, where instead of using single shooting, the same IPOPT formulation used in the diffusion framework was employed to solve the multiple-shooting problem. For the baseline, a random initial costate vector is sampled, the corresponding trajectory is propagated, and its discretized version is fed to IPOPT as the initial guess. 

A trial is considered successful if the maximum multiple-shooting continuity residual satisfies
$\max_k
\left\|
F(z_k,\Delta t_k)
-
z_{k+1}
\right\|_2
<
10^{-8}$,
where $F(\cdot)$ denotes one integration interval of the augmented state--costate dynamics. Table~\ref{tab:shift_generalization} summarizes the convergence statistics across all tested shifts.

\begin{figure}[t]
    \centering
    \includegraphics[width=1.0\linewidth]{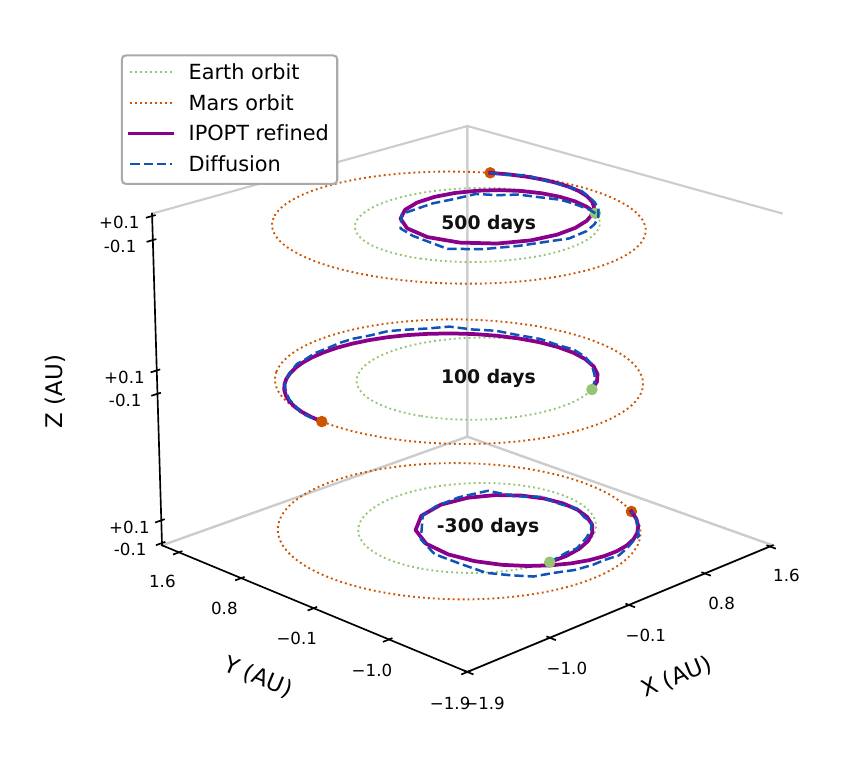}
    \\
    \caption{Representative converged trajectories for transfer windows shifted by $-300$, $+100$, and $+500$ days relative to the nominal departure epoch. AU denotes astronomical unit. }
    \label{fig:shift_examples}
\end{figure}

For all successful transfer windows, the proposed method achieved a $100\%$ convergence rate, whereas IPOPT initialized from random guesses failed to converge in some cases. The $+400$-shift is an exception in which IPOPT found more admissible trajectories; this specific trajectory is similar to the $+500$-shift trajectory (Fig.~\ref{fig:shift_examples}) with one revolution, but the engine operates for nearly the entire transfer. Several shifted transfer windows failed for both approaches. 
These regions correspond to combinations of spacecraft parameters and boundary conditions that either do not belong or are close to the boundary of the spacecraft reachable set \cite{bowerfind2024rapid}. All tests were run on a desktop computer with an Nvidia RTX 4090 GPU and an Intel i9-14900K CPU. On average, the diffusion pipeline (including refinement) took 18.5 seconds, while the baseline took 68.4 seconds. The baseline does not utilize GPU acceleration and only uses the CPU.

\begin{figure}[t]
    \centering
    \includegraphics[width=1.0\linewidth]{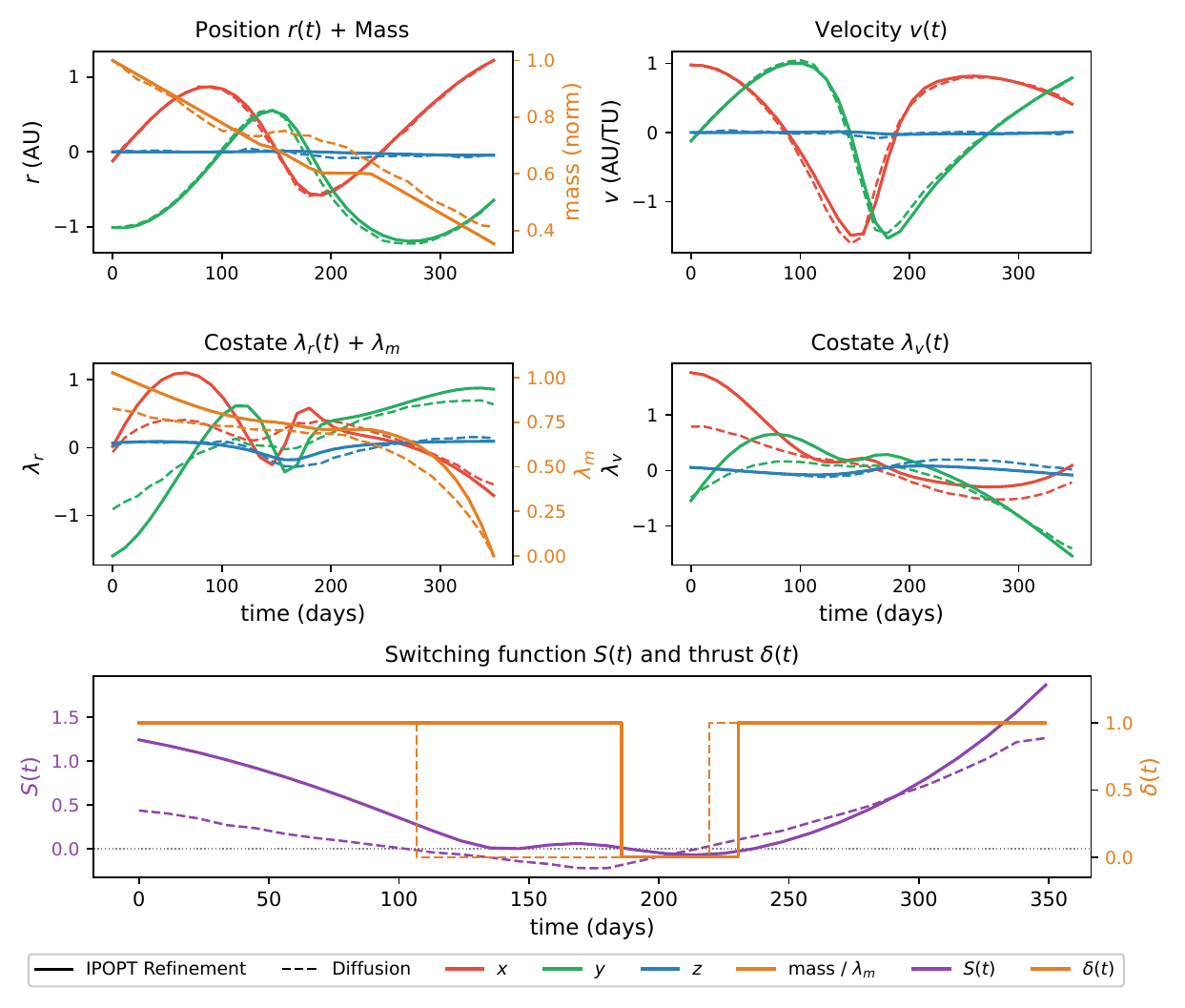}
    \caption{Comparison of states, costates, switching function and throttle plots before and after IPOPT refinement. The diffusion model provides a near admissible solution and IPOPT ensures full admissibility.}
    \label{fig:diffusion_refinement_state_costate}
\end{figure}

Three trajectories, corresponding to transfer windows shifted by $-300$, $+100$, and $+500$ days relative to the nominal Earth-to-Mars departure epoch, are shown in Fig.~\ref{fig:shift_examples}. Direct transfers and solutions with one orbital revolution are generated. Despite the substantial variation in planetary geometry, our model successfully generates initial guesses that can be refined into fully converged solutions. Moreover, in Fig.~\ref{fig:diffusion_refinement_state_costate}, the diffusion solution is compared against its IPOPT-refined version for the $-300$-day shift scenario.

\begin{table}[h]
\centering
\caption{Convergence rates over 250 trials and optimal final mass per departure-date shift.}
\label{tab:shift_generalization}
\footnotesize
\setlength{\tabcolsep}{3pt}
\renewcommand{\arraystretch}{1.05}
\begin{tabular}{c @{\hspace{4pt}{\color{black!25}\vline}\hspace{4pt}}
                c @{\hspace{4pt}{\color{black!25}\vline}\hspace{4pt}}
                c @{\hspace{4pt}{\color{black!25}\vline}\hspace{4pt}}
                r | c @{\hspace{4pt}{\color{black!25}\vline}\hspace{4pt}}
                    c @{\hspace{4pt}{\color{black!25}\vline}\hspace{4pt}}
                    c @{\hspace{4pt}{\color{black!25}\vline}\hspace{4pt}}
                    r}
\toprule
Shift & Ours & IPOPT & $m_f$ & Shift & Ours & IPOPT & $m_f$ \\
(d)   & (\%) & (\%)  & (kg)  & (d)   & (\%) & (\%)  & (kg)  \\
\midrule
$-700$ & $100.0$ & $57.6$ & $730.0$ & $+50$  & $100.0$ & $56.8$ & $659.6$ \\
$-600$ & $100.0$ & $48.8$ & $489.2$ & $+100$ & $100.0$ & $52.4$ & $722.7$ \\
$-500$ & $0.0$   & $0.0$  & ---     & $+200$ & $0.0$   & $0.0$  & ---     \\
$-400$ & $0.0$   & $0.0$  & ---     & $+300$ & $0.0$   & $0.0$  & ---     \\
$-300$ & $100.0$ & $18.4$ & $354.3$ & $+400$ & $0.4$   & $6.0$  & $285.0$ \\
$-200$ & $100.0$ & $31.2$ & $476.9$ & $+500$ & $100.0$ & $16.8$ & $393.5$ \\
$-100$ & $100.0$ & $54.4$ & $544.4$ & $+600$ & $100.0$ & $32.4$ & $450.1$ \\
$-50$  & $100.0$ & $59.2$ & $568.9$ & $+700$ & $100.0$ & $52.8$ & $486.4$ \\
$+0$   & $100.0$ & $62.0$ & $603.9$ &        &         &        &         \\
\bottomrule
\end{tabular}
\end{table}


\section{Conclusion}

This work introduced a diffusion-based framework for solving multiple-shooting indirect optimal control problems. To our knowledge, it is the first framework that integrates implicit one-step lag guidance with diffusion modeling and multiple-shooting methods to generate joint state--costate trajectories satisfying boundary conditions. A transformer architecture enables efficient handling of the resulting high-dimensional sequential representation and contextual one-step lag guidance, while a novel dataset-generation procedure provides diverse samples satisfying the optimality conditions. Results show improved convergence while preserving the optimality structure imposed by Pontryagin's Minimum Principle. The trained model also generalizes beyond the training dataset to Earth-to-Mars (7 states and 4 controls) direct-transfer trajectories, including one-revolution cases.

\bibliographystyle{IEEEtran}
\bibliography{bibliography}

\end{document}